\documentclass[11pt,a4paper]{article}
\pdfoutput=1

\usepackage[includeheadfoot,
            marginratio={1:1,2:3}, 
            width=412pt, 
            height=688pt,]{geometry}

\usepackage{amsmath}
\usepackage{amsfonts}
\usepackage{amssymb}
\usepackage{stmaryrd}
\usepackage{graphicx}
\usepackage{booktabs}

\usepackage{empheq}
\usepackage{paralist}
\usepackage{cite}
\usepackage[hidelinks]{hyperref}

\newcommand{\nc}{\newcommand}
\nc{\lb}{\llbracket}
\nc{\rb}{\rrbracket}
\nc{\gl}{\llbracket}
\nc{\gr}{\rrbracket}
\nc{\del}{\partial}

\newcommand{\eq}[1]{\begin{equation}
                     \begin{split} #1 \end{split}
                     \end{equation}}

\newcommand{\ov}{\overline}

\allowdisplaybreaks[2]
\numberwithin{equation}{section}
\begin{document}

\vspace*{-1.5cm}
\begin{flushright}
  {\small
  MPP-2019-110\\
  }
\end{flushright}

\vspace{1.5cm}
\begin{center}
{\LARGE
A Note on the dS Swampland Conjecture, \\[0.2cm]
 Non-BPS branes and K-theory\\[0.2cm]
} 
\vspace{0.4cm}

\end{center}

\vspace{0.35cm}
\begin{center}
  Ralph Blumenhagen$^{1}$, Max Brinkmann$^{1}$, Andriana Makridou$^{1,2}$
\end{center}

\vspace{0.1cm}
\begin{center} 
\emph{$^{1}$ Max-Planck-Institut f\"ur Physik (Werner-Heisenberg-Institut), \\ 
   F\"ohringer Ring 6,  80805 M\"unchen, Germany } \\[0.1cm] 
\vspace{0.25cm} 
\emph{$^{2}$ Ludwig-Maximilians-Universit{\"a}t M\"unchen, Fakult{\"a}t f{\"u}r Physik,\\ 
               Theresienstr.~37, 80333 M\"unchen, Germany}\\

\vspace{0.2cm}

 \vspace{0.3cm} 
\end{center} 

\vspace{0.5cm}

%%%%%%%%%%%%%%%%%%%%%%%%%%%%%%%%%%%%%%%%%%%%%%%
%%%%%%%%%%%%%%%%%%%%%%%%%%%%%%%%%%%%%%%%%%%%%%%
%%%%%%%%%%%%%%%%%%%%%%%%%%%%%%%%%%%%%%%%%%%%%%%
%%%%%%%%%%%%%%%%%%%%%%%%%%%%%%%%%%%%%%%%%%%%%%%
%%%%%%%%%%%%%%%%%%%%%%%%%%%%%%%%%%%%%%%%%%%%%%%
%%%%%%%%%%%%%%%%%%%%%%%%%%%%%%%%%%%%%%%%%%%%%%%
%%%%%%%%%%%%%%%%%%%%%%%%%%%%%%%%%%%%%%%%%%%%%%%
%%%%%%%%%%%%%%%%%%%%%%%%%%%%%%%%%%%%%%%%%%%%%%%

\begin{abstract}
We point out that the de Sitter swampland conjecture 
would  be falsified if classical fluxed Type IIA orientifold vacua
with a single non-BPS
D7-brane were  indeed part of the string theory landscape.
In other words, the dS swampland conjecture implies the 
cancellation of K-theory charges on a compact space.
\end{abstract}

\clearpage

%\tableofcontents

%%%%%%%%%%%%%%%%%%%%%%%%%%%%%%%%%%%%%%%%%%%%%%%
%%%%%%%%%%%%%%%%%%%%%%%%%%%%%%%%%%%%%%%%%%%%%%%
%%%%%%%%%%%%%%%%%%%%%%%%%%%%%%%%%%%%%%%%%%%%%%%
%%%%%%%%%%%%%%%%%%%%%%%%%%%%%%%%%%%%%%%%%%%%%%%
%%%%%%%%%%%%%%%%%%%%%%%%%%%%%%%%%%%%%%%%%%%%%%%
%%%%%%%%%%%%%%%%%%%%%%%%%%%%%%%%%%%%%%%%%%%%%%%
%%%%%%%%%%%%%%%%%%%%%%%%%%%%%%%%%%%%%%%%%%%%%%%
%%%%%%%%%%%%%%%%%%%%%%%%%%%%%%%%%%%%%%%%%%%%%%%

%\newpage

\section{Introduction}
\label{sec:intro}

Since its first proposal~\cite{Vafa:2005ui}, the swampland
program has recently been advanced to a set of intertwined 
swampland conjectures.
These are a set of quantitative properties
~\cite{ArkaniHamed:2006dz,Ooguri:2006in,Klaewer:2016kiy,Ooguri:2016pdq,Palti:2017elp,Obied:2018sgi,Cecotti:2018ufg,Garg:2018reu,Ooguri:2018wrx,Klaewer:2018yxi,Gonzalo:2019gjp,Heckman:2019bzm,Lust:2019zwm}
that a low-energy effective field theory should satisfy in order to admit a  UV completion
to a consistent theory of quantum gravity (see
~\cite{Palti:2019pca} for a recent review).
The main support for these conjectures derives from string theory, 
in particular from its landscape of compactifications.
Moreover, it was realized that the different swampland conjectures are
not unrelated, but rather form a tight web with many interrelations.
Thus they are mutually supporting each other and in this respect they 
are similar to the duality conjectures for the different perturbative
string theories, that were discovered in the nineties.

To better understand their origin and their consequences, these
swampland conjectures  need to be challenged by concrete string theory
constructions. For instance, the KKLT scenario~\cite{Kachru:2003aw} can still be considered
a challenge for the dS swampland conjecture, which forbids such dS
vacua in any theory of quantum gravity. Since this latter suspicion first arose
~\cite{Brennan:2017rbf,Danielsson:2018ztv},
the KKLT scenario has been scrutinized  in an ongoing debate.
One issue that arose was the existence of a 10D realization of KKLT
~\cite{Moritz:2017xto,Kallosh:2018wme,Akrami:2018ylq,Kallosh:2018psh,Hamada:2018qef,Kallosh:2019axr,Hamada:2019ack,Carta:2019rhx,Gautason:2019jwq},
while a second one is the question whether the effective field theory
in the warped throat is really controlled~\cite{Bena:2018fqc,Blumenhagen:2019qcg}.
Even though these arguments  have not yet
been completely settled, they revealed new aspects of the KKLT
scenario and their relation to the various swampland conjectures.

To reveal new connections  between the swampland conjectures and
fundamental aspects of string theory (that are often not derived 
but motivated by indirect arguments), one can also turn the logic
around and analyze what are the consequences for string theory, if
the swampland conjectures are assumed to be correct.
The  aim of this note is to provide an argument of this type that
involves the dS swampland conjecture and the cancellation of K-theory
charges for D-branes in string theory.
 
The dS swampland conjecture \cite{Obied:2018sgi} states that
$|\nabla V|\ge {c\over  M_{\rm pl}} \cdot V,\,$
where $c$ is of order one. The refined version of the conjecture~\cite{Garg:2018reu,Ooguri:2018wrx}
states that either the previous inequality or 
$min(\nabla_i\nabla_j V)\le -{c'\over  M_{\rm pl}^2} \cdot V$ has to hold, 
where $min(\nabla_i\nabla_j V)$ is the minimal eigenvalue of the Hessian matrix
and $c'$ is also of order one. 
Consistent with the dS no-go theorem of~\cite{Maldacena:2000mw},
these conjectures forbid (meta-)stable de Sitter vacua.  Such an inequality was first 
derived for classical flux vacua of Type IIA string theory in~\cite{Hertzberg:2007wc}. 
In this paper we will consider such Type IIA orientifolds and will try to challenge the
conjecture by introducing non-BPS branes in the background.
This sources  supersymmetry breaking and provides a positive
contribution to the scalar potential. 

Such branes are known to exist e.g. in the Type I superstring, where
the non-BPS $D0$-branes are S-dual to massive perturbative  heterotic
string states \cite{Bergman:1999kq}. Non-BPS branes only couple to closed string states from
the NS-NS sector. They do not carry any R-R charge, so  
there is no Bianchi identity whose integrated form 
can give rise to tadpole cancellation conditions. Thus there is no
K-theory analogue to the condition that the total R-R
charge on a compact space has to vanish.
For stable non-BPS branes on orientifolds, the open string tachyon is projected out not by a GSO projection
but by the world-sheet parity operation.  This can only happen for a
single brane, so a stack of two branes is unstable and decays.
Mathematically, such non-BPS branes are described by $\mathbb Z_2$
valued K-theory classes\cite{Witten:1998cd} so that a single such brane has a
topological obstruction to decay. The question arises whether also the
total K-theory charge on a compact space has to cancel, i.e. that it is even.
Arguments in favour of this could be indirectly derived for certain
cases by analyzing global anomalies on probe-branes \cite{Uranga:2000xp}.

In this note, we will find evidence for the following
\begin{quotation}
\noindent
{\bf Proposition:}
{\it If the (refined) de Sitter swampland conjecture is correct, then the K-theory
  charge on a compact space has to be trivial.}
\end{quotation}

\noindent
We will work in fluxed Type IIA orientifolds, where all closed string
moduli can be stabilized by NS-NS and R-R fluxes.
Via T-duality, one expects that such Type IIA orientifolds with intersecting $D6$-branes 
will also admit non-BPS branes. In section 2, after providing the salient features of
Type IIA orientifolds,  we determine which non-BPS branes are expected to exist. 

In section 3, assuming that K-theory charges are not
cancelled, we analyze  how the dS no-go theorem for fluxed Type IIA
orientifolds is affected by the presence of such branes. 
Indeed, we find that the dS no-go theorem does not go through.
Finally, we construct a supergravity toy model, which after
adding a single non-BPS brane, admits a dS vacuum. 
This provides strong evidence for  the proposition.

\section{Type IIA orientifolds with fluxes and branes}
\label{sec_two}

In this section we briefly review the set-up that are Type IIA
orientifolds with intersecting D6-branes and moduli stabilized by
NS-NS and R-R fluxes. In contrast to Type IIB, here all closed string moduli can be
stabilized already at string tree-level. Moreover, we argue via T-duality that in
these Type IIA orientifolds there should exist stable non-BPS branes.  

\subsection{Basics of Type IIA orientifolds}

We consider a Type IIA orientifold on a Calabi-Yau threefold. Here
the orientifold projection is given by $\Omega \ov\sigma (-1)^{F_L}$,
where $\Omega:(\tau,\sigma)\to (\tau,-\sigma)$ is the world-sheet
parity transformation, $\ov\sigma$ an anti-holomorphic involution
of the CY and $F_L$ the left-moving space-time fermion number
operator. The anti-holomorphic involution acts on the K\"ahler form $J$
and the covariantly constant holomorphic three-form $\Omega_3$ as
\eq{
           \ov\sigma:J\to -J\,,\qquad \ov\sigma:\Omega_3\to
           \ov\Omega_3\,.
}
The orientifold projection breaks the ${\cal N}=2$ supersymmetry of
Type IIA Calabi-Yau compactifications to ${\cal N}=1$.
The massless spectrum in the closed string sector of such a compactification is determined by
the equivariant cohomology groups  shown in table~\ref{table_closedspectrum}.

\begin{table}[ht]
\centering
  \begin{tabular}{ccc}
  \toprule
${\cal N}=1$ multiplet  &  {\rm State}  & {\rm Cohomology}    \\
\midrule
chiral &  $U=\int_{\Sigma_3} e^{-\phi} {\rm Re}(\Omega_3)+i \int_{\Sigma_3} C_3$  &
$\Sigma_3\in H_+^3(X) $    \\ \addlinespace
chiral & $T=\int_{\Sigma_2} J+ i \int_{\Sigma_2} B$  & $\Sigma_2\in H_-^2(X)$     \\ \addlinespace
vector &   $V=\int_{\Sigma_2} C_3$  & $\Sigma_2\in H_+^2(X)$    \\
\bottomrule
  \end{tabular}
  \caption{Massless  spectrum of Type IIA orientifold.}
  \label{table_closedspectrum}
\end{table}

\noindent
The universal chiral multiplet with $\Sigma_3\in H^{3,0}\oplus H^{0,3}$ 
is denoted as $S$. The fixed point locus of $\ov\sigma$ defines
the location of $O6$-planes, whose $C_7$-form tadpole needs to be
cancelled by the introduction of stacks of in general intersecting
$D6$-branes. Such intersecting $D6$-brane models have been studied in
detail in the past (see~\cite{Blumenhagen:2006ci,Ibanez:2012zz} for reviews and further references).

The kinetic terms for the moduli are encoded in the tree-level
K\"ahler potential
\eq{
           K=-\log \bigg(  {4\over 3} \int J\wedge J \wedge  J\bigg) 
%- \log (S+\ov S) 
- 2\log\bigg( \int {\rm Re}(\Omega_c)
           \wedge    \star {\rm Re}(\Omega_c)\bigg)\,
}
with  $\Omega_c=e^{-\phi} {\rm Re}(\Omega_3) +i C_3$.
 The $S,U,T$ moduli can (all) be fixed by the introduction of fluxes
 in Type IIA. Concretely, the complex structure moduli $U$ and $S$ can
 be fixed by NS-NS three-form flux $H=dB$, whereas the K\"ahler moduli $T$
receive a potential from non-vanishing R-R fluxes $F_0,F_2,F_4$ and $F_6$.
The corresponding superpotential is of Gukov-Vafa-Witten type
\eq{  
\label{superpot}
           W=i \int_X  \Omega_c \wedge H +
           \int_X  e^{i J_c} \wedge {\cal F}
 }
where ${\cal F}$ denotes the formal sum over all even R-R  fluxes and
$J_c=J+iB$ is the complexified K\"ahler form.
Explicit expressions can
be obtained by expanding all forms in the respective  cohomological
bases of $H^2_{\pm}$ and $H^3_{\pm}$. Here the orientifold even fluxes take values in
the equivariant cohomology groups listed in table~\ref{table_fluxcohom}.

\begin{table}[ht]
\centering
  \begin{tabular}{c c}
  \toprule
  Flux    & {\rm Cohomology}    \\
    \midrule
 &  \\[-0.4cm]
$H$ & $H_-^3(X)$  \\[0.1cm]
$\{ F_0, F_2,F_4 ,F_6\}$ & $\{H^0_+, H^2_-, H^4_+, H^6_-    \}$ \\
\bottomrule
  \end{tabular}
  \caption{Cohomology groups of orientifold even fluxes.}
  \label{table_fluxcohom}
\end{table}

Noting that the total volume form ${\rm vol}=J^3$ is $\ov\sigma$-odd, these
fluxes are precisely those that can give a non-vanishing contribution
to the superpotential~\eqref{superpot}.

\subsection{Non-BPS branes}

It is known that due to the orientifold projection there can exist 
stable non-BPS branes (see for instance the review~\cite{Sen:1999mg}) that correspond
to torsional K-theory groups~\cite{Witten:1998cd}.
The simplest example is the Type I superstring that besides
BPS $D1$, $D5$ and $D9$ branes also contains the stable non-BPS branes
listed in table \ref{table_nonBPS}~\cite{Frau:1999qs}. The boundary states of these branes only
contain contributions from the NS-NS sector. Thus in the  loop-channel
annulus amplitude the GSO projection is missing, and no R-R tadpole conditions 
for these branes arise.

The tachyon in the open string NS-sector is instead projected  out by the 
orientifold projection $\Omega$. This only works for a single such 
brane, hence two non-BPS branes on top of each other are unstable. 
In other words, the corresponding K-theory group is $\mathbb Z_2$ valued.
\begin{table}[ht]
\centering
  \begin{tabular}{c c c}
  \toprule
   {\rm non-BPS brane}  & {\rm Tension}   & {\rm K-theory}  \\
    \midrule
& & \\[-0.4cm]
 $\widehat D8$  & $\sqrt 2\, T_{D8}$   & $KO(S^1)=\mathbb Z_2$  \\
$\widehat D7$  & $2\, T_{D7} $   & $KO(S^2)=\mathbb Z_2$  \\
  $\widehat D0$  & $\sqrt 2\, T_{D0}  $   & $KO(S^9)=\mathbb Z_2$  \\
$\widehat D({\rm -1})$  & $2\, T_{D(-1)}$   & $KO(S^{10})=\mathbb Z_2$ \\
\bottomrule
  \end{tabular}
  \caption{Stable non-BPS branes for the  Type I superstring.}
  \label{table_nonBPS}
\end{table}
The lower dimensional $\widehat D0$ and $\widehat D(-1)$ branes are not
of interest to us as they do not fill the entire four-dimensional
Minkowski space. 

To get an idea of how this story generalizes to Type IIA orientifolds
on  Calabi-Yau threefolds, we consider a $T^6=(T^2)^3$ compactification
of the Type I string and apply T-duality along the three $y_i$
directions so that the anti-holomorphic involution becomes $z_i\to \ov
z_i$ with $z_i=x_i+ i y_i$.  Then for instance a $\widehat D8$ brane that is
localized at $y_3=0$ is mapped by T-duality to a $\widehat D 7$ brane wrapping the
four-cycle spanned by  $(x_1,x_2,x_3,y_3)$. Therefore, under  $\ov \sigma$ 
this four-cycle $\Sigma_4$ transforms as $\Sigma_4\to -\Sigma_4$ so it is in 
the $H_4^-(X)$ homology.

If the initial $\widehat D8$ brane is instead
localized  at $x_3=0$, after T-duality one obtains a $\widehat D5$ brane 
 wrapping the two-cycle spanned by  $(x_1,x_2)$. In principle the $\widehat
 D8$ brane could also wrap a general $(p,q)$ one-cycle on the third
 $T^2$, in which case one gets a $\widehat D 7$ brane  equipped with a non-trivial
line-bundle. In this paper, we are not considering such fluxed non-BPS
branes,  mainly as the simpler non-fluxed ones are sufficient for our
purpose. To get the full spectrum of stable non-BPS branes one has to
determine the corresponding K-theory groups, which is however not an
easy task.

Applying T-duality to the non-BPS $\widehat D8$
and $\widehat D7$ branes in all possible space-time filling positions
we find the (non-fluxed) non-BPS
branes for Type IIA orientifolds listed in table \ref{table_nonBPSIIA}.
\begin{table}[ht]
\centering
  \begin{tabular}{c c c}
  \toprule
   {\rm Type I}  & {Type IIA}   & {Homology}  \\
    \midrule
& & \\[-0.4cm]
 $\widehat D8$  & $\widehat D7$   & $H_4^-(X)$  \\
     & $\widehat D5$   & $H_2^+(X)$  \\[0.1cm]
\hline
& & \\[-0.4cm]
$\widehat D7$  & $\widehat D4$   & $H_1^+(X)$  \\
     & $\widehat D6$   & $H_3^-(X)$  \\
     & $\widehat D8$   & $H_5^+(X)$  \\
     \bottomrule
  \end{tabular}
  \caption{Stable (non-fluxed) non-BPS branes for the  Type IIA orientifolds.}
  \label{table_nonBPSIIA}
\end{table}
Note that on a Calabi-Yau $\widehat D4$ and $\widehat D8$ are not
supported, as there are no homological one- and five-cycles. Moreover, 
$\widehat D6$ branes\footnote{As shown in \cite{MarchesanoBuznego:2003axu}, also the BPS $D6$
  branes contribute to the  K-theory charge carried  by
  the non-BPS  $\widehat D6$ branes. The vanishing of the global
  Witten-anomaly on a probe D6-brane of $SP$-type implied that 
   this K-theory charge has to be trivial.}
 are in danger of developing a Freed-Witten
anomaly, as the $H$-flux is also supported in the odd-(co-)homology
group\footnote{This effect is taken care of by computing the
  H-twisted K-theory groups.}.

On the other hand the $\widehat D5$ brane is free from Freed-Witten
anomalies while  a $\widehat D7$ brane can be made safe 
by wrapping it on a four-cycle that does not contain any homological three-cycle, 
like e.g. a del-Pezzo surface. 
Moreover, a del-Pezzo surface is rigid so
there are no transversal deformations of the non-BPS brane. By
wrapping the intersecting $D6$-branes also on rigid three-cycles  in
$H_3^{+}(X)$, one can avoid that the non-BPS branes and the $D6$-branes
can come close to each other.  Their mutual attraction or repulsion
will only lead to a one-loop potential for the closed string moduli
that in the perturbative regime is  expected to be subdominant
relative to the tree-level flux induced potential.

As we have seen, the prime candidates for our purpose are the non-BPS 
$\widehat D5$ and $\widehat D7$ branes (wrapped on rigid four-cycles) in Type IIA orientifolds.
Note that they are not directly coupled to  the complexified K\"ahler
moduli, that belong	 to the $H_2^+(X)$ homology. However,
the existence of such non-BPS branes comes along with extra $U(1)$
symmetries from the closed string R-R sector. In addition, while the
tachyon is projected out, 
 a single such non-BPS brane will support an open string $U(1)$ gauge
 field on its world-volume. 
%In principle there can be a whole dynamical
% hidden sector localized on a set of space-time filling non-BPS branes.

\section{The failure of the dS no-go theorem} 

In this section we analyze the scalar potential of fluxed Type IIA
orientifolds with a single non-BPS brane included in the background. First, 
we review the no-go theorem for dS minima in Type IIA and
investigate what happens if the tension of the non-BPS branes
is included.

\subsection{Analysis of  no-go theorem with non-BPS branes}\label{nogo}

In \cite{Hertzberg:2007wc} it was shown that classical Type IIA flux
vacua on a Calabi-Yau space with D6-branes and
O6-planes satisfy  a lower bound for the slow-roll parameter $\epsilon$,
namely
\eq{
                   {M_{\rm pl}^2\over 2} \left({\nabla V\over V}\right)^2 \ge {27\over 13}
}     
for $V>0$. Clearly, this forbids dS vacua. Let us redo their
computation while also allowing for the presence of non-BPS $\widehat D5$ and $\widehat D7$ branes. 
The idea of \cite{Hertzberg:2007wc} is to keep track of the contributions to the
scalar potential from the individual constituents, e.g. fluxes and
D-branes, and in particular tracking how the individual contributions scale with respect to the universal 
modulus $s={\rm Re}(S)=e^{-\phi} {\rm vol}^{1/2}$ and the overall volume modulus $t={\rm
  vol}^{1/3}$. Here the convention is that the volume is measured in
string frame and eventually one transforms the 4D metric $g_s$ to Einstein
frame via $g_s=e^{2\phi} g_E$. As a consequence, the string scale and
  the 4D Planck scale are related as $M_{\rm pl}\sim M_{s} {\rm vol}^{1/2}$.

In this way, one finds the following form of the total potential
\eq{
\label{potscaling}
  V={A_H\over s^2 t^3}+\sum_{p\,{\rm even}} {A_{F_p}\over s^4 t^{p-3}} +{A_{D6}\over
    s^3}-{A_{O6}\over s^3 } +{A_{\widehat D 5}\over s^3 t^{1\over 2}}
  +{A_{\widehat D 7} t^{1\over 2}\over s^3 }
}
where we have indicated the different contributions from fluxes,
D-branes and O-planes.
The coefficients in the numerators are positive semi-definite and in general are complicated functions of all the
other present moduli. Next, one computes the combination
\eq{
\label{typetwonogo}
       t {\partial V\over \partial t}+  3\, s 
   {\partial  V\over \partial s} = -9 V - \sum_p {p A_{F_p}\over s^4 t}
   -{1\over 2} {A_{\widehat D 5}\over s^3 t^{1\over 2}}+{1\over 2}{A_{\widehat D 7} t^{1\over 2}\over s^3 }\,.
}
For $A_{\widehat D 7}=0$ this implies that any extremum $\partial_t V=\partial_s V=0$ will
satisfy $V\le 0$. However, due to the positive sign in front of the
last term in \eqref{typetwonogo}, this dS no-go argument does not go through when 
there is a  non-BPS $\widehat D 7$ brane present in the
background\footnote{That the no-go does not apply for $Dp$ branes with
  $p>6$ was already observed in \cite{Hertzberg:2007wc}, though not applied to non-BPS
  branes.}.
Thus, there still exists a chance that Type IIA orientifolds with a non-BPS
$\widehat D 7$ brane admit dS minima. 

\subsection{A simple SUGRA example}\label{SUGRAex}

In this section, we analyze whether a simple SUGRA model does indeed
lead to dS vacua when including a non-BPS $\widehat D 7$ brane.
This model has only two moduli, a K\"ahler modulus
$T$ and the universal modulus $S$. One can think of it as a toroidal
$STU$-model where we have identified $T=T_1=T_2=T_3$ and $S=U_1=U_2=U_3$.
The K\"ahler potential becomes
\eq{     
               K=-3\log (T+\ov T) -4 \log(S+\ov S)
}
and the flux induced superpotential
\eq{
           W=-4 ihS + f_6 + 3 i f_4 T - 3 f_2 T^2 -i f_0 T^3\,.
}
If this model arises as an effective one after integrating out all
other moduli, the coefficients $h,f_i$ are not necessarily quantized
integers. Nevertheless, here we treat them as integers to see how
much freedom we have in ``tuning'' such a model even if only 
integers are involved.

To further simplify the setting, we choose $f_6=f_2=0$. The resulting
supergravity scalar potential is minimized for vanishing axions 
${\rm Im}(S)={\rm Im}(T)=0$, hence the potential for the saxions $s$
and $t$ becomes
\eq{\label{noD7pot}
                 V_F= {h^2\over 8\, s^2 t^3}+ {f_0^2\, t^3\over 32\, s^4 }
                 + {3 f_4^2 \over 32\, s^4 t}- {f_0 h\over  4\, s^3} 
} 
which features precisely the scaling expected in \eqref{potscaling}. Note that
the last term scales like a D6-brane contribution. This makes sense as
the flux combination $h f_0$ contributes to the D6-brane tadpole
cancellation condition, which has implicitly been taken into account
when expressing the scalar potential as an F-term of a SUGRA model.

Minimizing the potential with respect to $s, t$ one obtains a
supersymmetric AdS-minimum at
\eq{
            s_0= {\textstyle \sqrt{20\over 27}}\, {f_4^{3\over 2}\over f_0^{1\over 2} h}\,,\qquad
            t_0 = {\textstyle\sqrt{5\over 3}}\, {f_4^{1\over 2} \over f_0^{1\over 2}}\,,\qquad
                    V_0\approx - 0.059\, {f_0^{5\over 2} h^4\over
                      f_4^{9\over 2}} M_{\rm pl}^4\,.
} 
This model is of the flux-scaling type promoted in  \cite{Blumenhagen:2015kja}.
To be in the perturbative regime we require $t_0\gg 1$ and
$e^{-\phi_0}=s_0/t_0^{3/2}\sim (f_0 f_4^3)^{1/4}/h\gg 1$.  This can be
achieved by choosing the flux $f_4$ sufficiently large.

Now we add a contribution from a single non-BPS $\widehat D 7$ brane
\eq{
\label{totalpot}
                 V=V_F + {A_{\widehat D 7}\, t^{1\over 2}\over s^3 }
}
where the coefficient $A_{\widehat D 7}$ is a fixed number, that can
be determined from a dimensional reduction of the Type IIA action as
$A_{\widehat D 7}=\sqrt 2\cdot 3/16$.
% for definiteness\footnote{
%The precise value of this constant is arbitrary, since any different 
%constant can be identified with our choice by rescaling the fluxes.
%}.
In order to see whether dS minima are possible, we first treat
$A_{\widehat D 7}$ as a free parameter and try to solve for Minkowski
minima. Thus, solving $\partial_s V=\partial_t V=V=0$ for the three
variables $\{ s,t,A_{\widehat D 7}\}$ we find
\eq{
\label{minkomin}
         s_0\approx  3.64\, {f_4^{3\over 2}\over f_0^{1\over 2} h}\,,
         \qquad 
       t_0\approx 1.87 {f_4^{1\over 2} \over
         f_0^{1\over 2}} \,,\qquad  A_{\widehat D 7}\approx 0.081 {f_0^{5\over 2} h\over f_4^{1\over 4}}\,.
}
If we tried to do the same for the contribution of a non-BPS  $\widehat D 5$
brane, we would  not find any solution with all $s,t,A_{\widehat D 5}$
coming out positively. This case is consistent with the above no-go theorem.

\begin{figure}[tbh]
  \centering
  \includegraphics[width=0.6\textwidth]{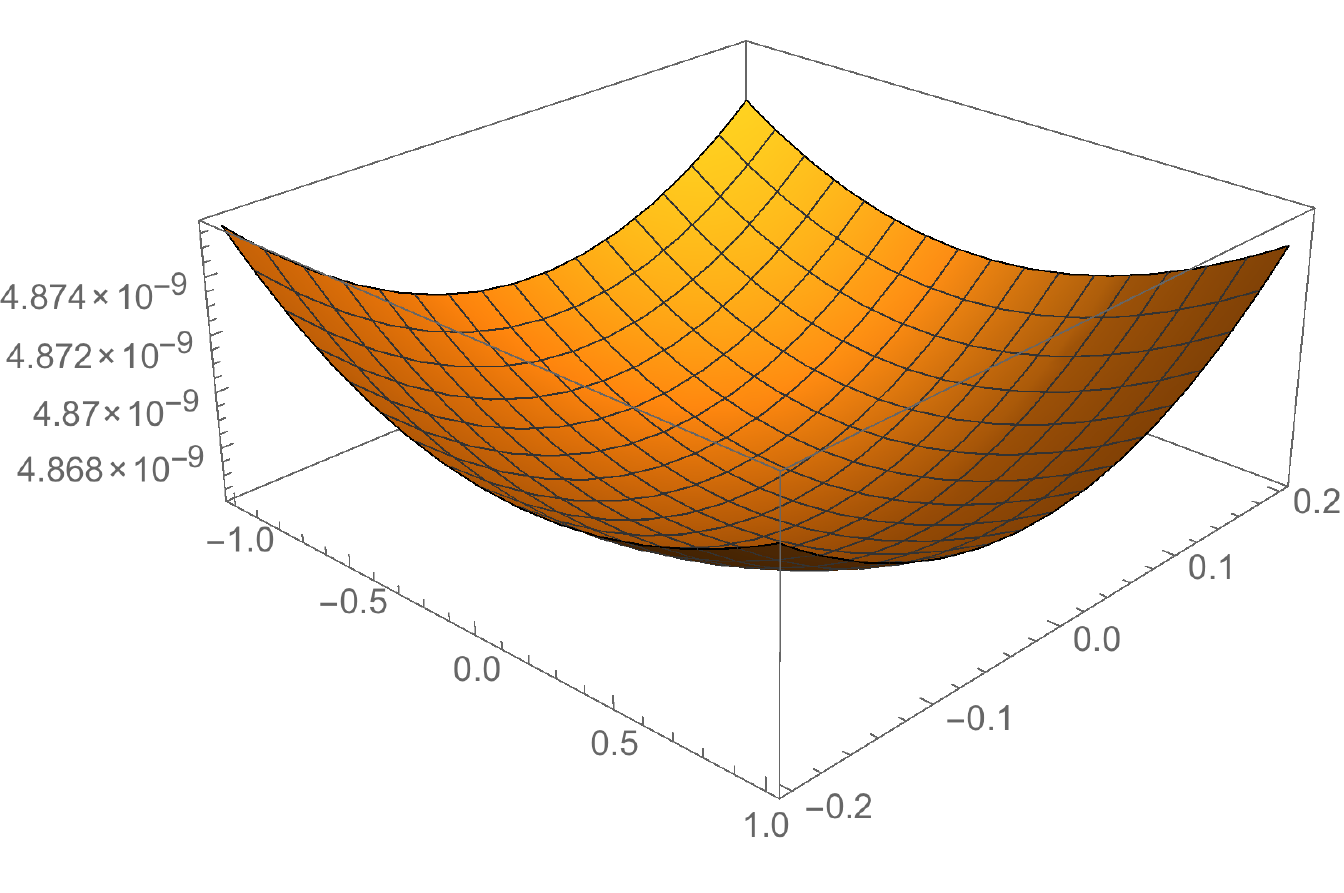}
  \begin{picture}(0,0)
%  \put(-90, 5){\footnotesize$x$}
%   \put(0,80){\footnotesize$y$}
\end{picture}
  \caption{Plot of the potential $V\big(s_0+x+y,t_0+0.02 (x-y)\big) $
    with $h=3, f_0=2, f_4=23$ in the range $|x|\le 1$
    and  $|y|\le 0.2$.}
  \label{fig:dSvacuum}
\end{figure}

To obtain a dS minimum, we must now find integers $h,f_0,f_4$ so that $A_{\widehat D 7}$
in \eqref{minkomin} is slightly smaller than the true value $A_{\widehat D 7}=\sqrt
2\cdot 3/16$. One choice is $h=3, f_0=2, f_4=23$. Plugging these values into
the scalar potential \eqref{totalpot} with $A_{\widehat D
  7}=\sqrt 2 \cdot 3/16$, a numerical analysis shows that the resulting model features a dS
minimum at $s_0\approx 105.17$ and $t_0\approx 6.54$ with $V_0=4.87\cdot
10^{-9} M_{\rm pl}^4$.  The form of the potential around this minimum
is shown in figure \ref{fig:dSvacuum}. It is evident that we have indeed found a minimum
and not a saddle point, which is also verified by the fact that 
$min(\nabla_i\nabla_j V)|_{(s_0,t_0)}\approx 3.711 \cdot 10^{-8} M_{\rm pl}^2$.

%%%%%%%%%%%%
%%%%%%%%%%%%

\subsection{Discussion}

Thus, this simple SUGRA model indicates that there should 
exist dS minima in Type IIA orientifolds with a single non-BPS $\widehat D 7$
brane in the background. This is not yet a full-fledged
string theory compactification, as we have only considered a stringy model with two moduli 
via suitable identifications in the STU-model and then added a hard supersymmetry
breaking sector to the theory by hand. 

We can however provide two arguments supporting the assumption 
that such a set-up is likely to be controlled. 
Since the tension of the non-BPS brane scales in the same way
as the tension of D-branes, we expect that their backreaction on the
geometry is controlled by the string coupling constant and therefore
should be small  in the weak coupling regime $g_s\ll 1$. Second, once we make
sure that the branes do not have open string moduli,
we also expect the attraction/repulsion with  the other $D6$-branes to 
only lead to $g_s$ suppressed subleading contributions to the tree-level scalar
potential. 

Having such stable dS Sitter minima is in conflict with the dS
swampland conjecture. It is easy to see that even the less constraining refined dS conjecture
does not provide a way out, since both the proposed inequalities are violated\footnote{In more 
realistic compactifications that feature a larger number of moduli, 
it would indeed be important to check the refined version of the conjecture to make sure that 
the dS vacuum does not arise only due to ignoring unstable directions in the moduli space.}. 
Instead of claiming that we found a counter example, we would rather consider this as 
evidence that something is wrong with our string theory set-up.
The obvious candidate here is that a single  non-BPS $\widehat D 7$ on a
compact transverse space should not be allowed.
But this will be precisely guaranteed if the K-theory charge is cancelled
(mod 2).

In this case there must be an even number of non-BPS $\widehat D 7$ branes
wrapping a four-cycle in $H_4^-(X)$. Note that the intersecting BPS
$D6$-branes do not contribute to this charge. An even number of 
non-BPS $\widehat D 7$ branes will however be unstable and decay.
In the case where two such branes do not carry any additional line
bundle, they can completely annihilate. If one of them is equipped
with an additional line bundle, we expect that open string tachyons will
appear, connecting the two branes. In this case, we expect that the two
non-BPS $\widehat D 7$ branes will decay to non-BPS $\widehat D 5$ branes
that might further decay. In any case, the  $\widehat D 7$ brane
contribution \eqref{totalpot} to the scalar potential will disappear
after the decay and along with it the dS minimum.

Therefore, we found evidence for our proposition that the dS swampland conjecture 
implies that K-theory charges should be cancelled on compact spaces.

\section{Conclusions}\label{sec_concl}

In this note we have pointed out that Type IIA orientifolds would
support dS minima if single non-BPS $\widehat D 7$ branes could be consistently
added to the background. First, via T-duality we argued which 
non-BPS branes are expected to be present in Type IIA orientifolds
and how they are related to equivariant homology classes. It turned
out that those homology classes that appear are not related to fluxes 
and moduli fields. 

We then investigated how the dS no-go theorem of \cite{Hertzberg:2007wc} is affected once one 
also takes the positive contribution of a  non-BPS brane  to the scalar potential into account. 
While for non-BPS $\widehat D 5$ branes the no-go still applies, for 
non-BPS $\widehat D 7$ branes it does not. We were able to provide a supergravity toy model
which after slight tuning of the fluxes indeed features a dS minimum.
The dS swampland conjecture can then only hold if the inclusion of single non-BPS branes is 
inherently forbidden. This exactly implies cancellation of K-theory charges.

Since the net of swampland conjectures is tightly knit, it could be interesting to 
search for further evidence for the cancellation of K-theory charges 
from other conjectures. A natural candidate to check would be the recently 
proposed Strong Scalar Weak Gravity Conjecture~\cite{Gonzalo:2019gjp}, once its generalization
to multiple scalar fields is clear.

\vspace{0.5cm}

\noindent
\subsubsection*{Acknowledgments}
We would like to thank Luis Ib\'a\~nez, Daniel Junghans and Daniel Kl\"awer for discussions.

\vspace{1cm}
%\clearpage
%\appendix

%%%%%%%%%%%%%%%%%%%%%%%%%%%%%%%%%%%%%%%%%%%%%%%
%%%%%%%%%%%%%%%%%%%%%%%%%%%%%%%%%%%%%%%%%%%%%%%
%%%%%%%%%%%%%%%%%%%%%%%%%%%%%%%%%%%%%%%%%%%%%%%
%%%%%%%%%%%%%%%%%%%%%%%%%%%%%%%%%%%%%%%%%%%%%%%
%%%%%%%%%%%%%%%%%%%%%%%%%%%%%%%%%%%%%%%%%%%%%%%
%%%%%%%%%%%%%%%%%%%%%%%%%%%%%%%%%%%%%%%%%%%%%%%  
%%%%%%%%%%%%%%%%%%%%%%%%%%%%%%%%%%%%%%%%%%%%%%%
%%%%%%%%%%%%%%%%%%%%%%%%%%%%%%%%%%%%%%%%%%%%%%%

%%%%%%%%%%%%%%%%%%%%%%%%%%%%%%%%%%%%%%%%%%%%%%%
%%%%%%%%%%%%%%%%%%%%%%%%%%%%%%%%%%%%%%%%%%%%%%%
%%%%%%%%%%%%%%%%%%%%%%%%%%%%%%%%%%%%%%%%%%%%%%%
%%%%%%%%%%%%%%%%%%%%%%%%%%%%%%%%%%%%%%%%%%%%%%%

%\clearpage
\bibliography{references}  

\providecommand{\href}[2]{#2}\begingroup\raggedright\begin{thebibliography}{10}

\bibitem{Vafa:2005ui}
C.~Vafa, ``{The String landscape and the swampland},''
  \href{http://arxiv.org/abs/hep-th/0509212}{{\ttfamily arXiv:hep-th/0509212
  [hep-th]}}.

\bibitem{ArkaniHamed:2006dz}
N.~Arkani-Hamed, L.~Motl, A.~Nicolis, and C.~Vafa, ``{The String landscape,
  black holes and gravity as the weakest force},''
  \href{http://dx.doi.org/10.1088/1126-6708/2007/06/060}{{\em JHEP} {\bfseries
  06} (2007) 060}, \href{http://arxiv.org/abs/hep-th/0601001}{{\ttfamily
  arXiv:hep-th/0601001 [hep-th]}}.

\bibitem{Ooguri:2006in}
H.~Ooguri and C.~Vafa, ``{On the Geometry of the String Landscape and the
  Swampland},'' \href{http://dx.doi.org/10.1016/j.nuclphysb.2006.10.033}{{\em
  Nucl. Phys.} {\bfseries B766} (2007) 21--33},
  \href{http://arxiv.org/abs/hep-th/0605264}{{\ttfamily arXiv:hep-th/0605264
  [hep-th]}}.

\bibitem{Klaewer:2016kiy}
D.~Kl{\"a}wer and E.~Palti, ``{Super-Planckian Spatial Field Variations and
  Quantum Gravity},'' \href{http://dx.doi.org/10.1007/JHEP01(2017)088}{{\em
  JHEP} {\bfseries 01} (2017) 088},
  \href{http://arxiv.org/abs/1610.00010}{{\ttfamily arXiv:1610.00010
  [hep-th]}}.

\bibitem{Ooguri:2016pdq}
H.~Ooguri and C.~Vafa, ``{Non-supersymmetric AdS and the Swampland},''
  \href{http://dx.doi.org/10.4310/ATMP.2017.v21.n7.a8}{{\em Adv. Theor. Math.
  Phys.} {\bfseries 21} (2017) 1787--1801},
  \href{http://arxiv.org/abs/1610.01533}{{\ttfamily arXiv:1610.01533
  [hep-th]}}.

\bibitem{Palti:2017elp}
E.~Palti, ``{The Weak Gravity Conjecture and Scalar Fields},''
  \href{http://dx.doi.org/10.1007/JHEP08(2017)034}{{\em JHEP} {\bfseries 08}
  (2017) 034}, \href{http://arxiv.org/abs/1705.04328}{{\ttfamily
  arXiv:1705.04328 [hep-th]}}.

\bibitem{Obied:2018sgi}
G.~Obied, H.~Ooguri, L.~Spodyneiko, and C.~Vafa, ``{De Sitter Space and the
  Swampland},'' \href{http://arxiv.org/abs/1806.08362}{{\ttfamily
  arXiv:1806.08362 [hep-th]}}.

\bibitem{Cecotti:2018ufg}
S.~Cecotti and C.~Vafa, ``{Theta-problem and the String Swampland},''
  \href{http://arxiv.org/abs/1808.03483}{{\ttfamily arXiv:1808.03483
  [hep-th]}}.

\bibitem{Garg:2018reu}
S.~K. Garg and C.~Krishnan, ``{Bounds on Slow Roll and the de Sitter
  Swampland},''
\href{http://arxiv.org/abs/1807.05193}{{\ttfamily arXiv:1807.05193 [hep-th]}}.
%%CITATION = ARXIV:1807.05193;%%.

\bibitem{Ooguri:2018wrx}
H.~Ooguri, E.~Palti, G.~Shiu, and C.~Vafa, ``{Distance and de Sitter
  Conjectures on the Swampland},''
  \href{http://dx.doi.org/10.1016/j.physletb.2018.11.018}{{\em Phys. Lett.}
  {\bfseries B788} (2019) 180--184},
  \href{http://arxiv.org/abs/1810.05506}{{\ttfamily arXiv:1810.05506
  [hep-th]}}.

\bibitem{Klaewer:2018yxi}
D.~Kl{\"a}wer, D.~L{\"u}st, and E.~Palti, ``{A Spin-2 Conjecture on the
  Swampland},'' \href{http://dx.doi.org/10.1002/prop.201800102}{{\em Fortsch.
  Phys.} {\bfseries 67} no.~1-2, (2019) 1800102},
  \href{http://arxiv.org/abs/1811.07908}{{\ttfamily arXiv:1811.07908
  [hep-th]}}.

\bibitem{Gonzalo:2019gjp}
E.~Gonzalo and L.~E. Ib{\'a}{\~n}ez, ``{A Strong Scalar Weak Gravity Conjecture
  and Some Implications},'' \href{http://arxiv.org/abs/1903.08878}{{\ttfamily
  arXiv:1903.08878 [hep-th]}}.

\bibitem{Heckman:2019bzm}
J.~J. Heckman and C.~Vafa, ``{Fine Tuning, Sequestering, and the Swampland},''
  \href{http://arxiv.org/abs/1905.06342}{{\ttfamily arXiv:1905.06342
  [hep-th]}}.

\bibitem{Lust:2019zwm}
D.~L{\"u}st, E.~Palti, and C.~Vafa, ``{AdS and the Swampland},''
\href{http://arxiv.org/abs/1906.05225}{{\ttfamily arXiv:1906.05225 [hep-th]}}.
%%CITATION = ARXIV:1906.05225;%%.

\bibitem{Palti:2019pca}
E.~Palti, ``{The Swampland: Introduction and Review},''
\newblock 2019.
\newblock \href{http://arxiv.org/abs/1903.06239}{{\ttfamily arXiv:1903.06239
  [hep-th]}}.

\bibitem{Kachru:2003aw}
S.~Kachru, R.~Kallosh, A.~D. Linde, and S.~P. Trivedi, ``{De Sitter vacua in
  string theory},'' \href{http://dx.doi.org/10.1103/PhysRevD.68.046005}{{\em
  Phys. Rev.} {\bfseries D68} (2003) 046005},
  \href{http://arxiv.org/abs/hep-th/0301240}{{\ttfamily arXiv:hep-th/0301240
  [hep-th]}}.

\bibitem{Brennan:2017rbf}
T.~D. Brennan, F.~Carta, and C.~Vafa, ``{The String Landscape, the Swampland,
  and the Missing Corner},'' \href{http://dx.doi.org/10.22323/1.305.0015}{{\em
  PoS} {\bfseries TASI2017} (2017) 015},
  \href{http://arxiv.org/abs/1711.00864}{{\ttfamily arXiv:1711.00864
  [hep-th]}}.

\bibitem{Danielsson:2018ztv}
U.~H. Danielsson and T.~{Van Riet}, ``{What if string theory has no de Sitter
  vacua?},'' \href{http://dx.doi.org/10.1142/S0218271818300070}{{\em Int. J.
  Mod. Phys.} {\bfseries D27} no.~12, (2018) 1830007},
  \href{http://arxiv.org/abs/1804.01120}{{\ttfamily arXiv:1804.01120
  [hep-th]}}.

\bibitem{Moritz:2017xto}
J.~Moritz, A.~Retolaza, and A.~Westphal, ``{Toward de Sitter space from ten
  dimensions},'' \href{http://dx.doi.org/10.1103/PhysRevD.97.046010}{{\em Phys.
  Rev.} {\bfseries D97} no.~4, (2018) 046010},
  \href{http://arxiv.org/abs/1707.08678}{{\ttfamily arXiv:1707.08678
  [hep-th]}}.

\bibitem{Kallosh:2018wme}
R.~Kallosh, A.~Linde, E.~McDonough, and M.~Scalisi, ``{de Sitter Vacua with a
  Nilpotent Superfield},'' \href{http://dx.doi.org/10.1002/prop.201800068}{{\em
  Fortsch. Phys.} {\bfseries 2018} (2018) 1800068},
  \href{http://arxiv.org/abs/1808.09428}{{\ttfamily arXiv:1808.09428
  [hep-th]}}.

\bibitem{Akrami:2018ylq}
Y.~Akrami, R.~Kallosh, A.~Linde, and V.~Vardanyan, ``{The Landscape, the
  Swampland and the Era of Precision Cosmology},''
  \href{http://dx.doi.org/10.1002/prop.201800075}{{\em Fortsch. Phys.}
  {\bfseries 67} no.~1-2, (2019) 1800075},
  \href{http://arxiv.org/abs/1808.09440}{{\ttfamily arXiv:1808.09440
  [hep-th]}}.

\bibitem{Kallosh:2018psh}
R.~Kallosh, A.~Linde, E.~McDonough, and M.~Scalisi, ``{4D models of de Sitter
  uplift},'' \href{http://dx.doi.org/10.1103/PhysRevD.99.046006}{{\em Phys.
  Rev.} {\bfseries D99} no.~4, (2019) 046006},
  \href{http://arxiv.org/abs/1809.09018}{{\ttfamily arXiv:1809.09018
  [hep-th]}}.

\bibitem{Hamada:2018qef}
Y.~Hamada, A.~Hebecker, G.~Shiu, and P.~Soler, ``{On brane gaugino condensates
  in 10d},'' \href{http://dx.doi.org/10.1007/JHEP04(2019)008}{{\em JHEP}
  {\bfseries 04} (2019) 008},
\href{http://arxiv.org/abs/1812.06097}{{\ttfamily arXiv:1812.06097 [hep-th]}}.
%%CITATION = ARXIV:1812.06097;%%.

\bibitem{Kallosh:2019axr}
R.~Kallosh, A.~Linde, E.~McDonough, and M.~Scalisi, ``{dS Vacua and the
  Swampland},'' \href{http://dx.doi.org/10.1007/JHEP03(2019)134}{{\em JHEP}
  {\bfseries 03} (2019) 134},
\href{http://arxiv.org/abs/1901.02022}{{\ttfamily arXiv:1901.02022 [hep-th]}}.
%%CITATION = ARXIV:1901.02022;%%.

\bibitem{Hamada:2019ack}
Y.~Hamada, A.~Hebecker, G.~Shiu, and P.~Soler, ``{Understanding KKLT from a 10d
  perspective},'' \href{http://dx.doi.org/10.1007/JHEP06(2019)019}{{\em JHEP}
  {\bfseries 06} (2019) 019},
\href{http://arxiv.org/abs/1902.01410}{{\ttfamily arXiv:1902.01410 [hep-th]}}.
%%CITATION = ARXIV:1902.01410;%%.

\bibitem{Carta:2019rhx}
F.~Carta, J.~Moritz, and A.~Westphal, ``{Gaugino condensation and small uplifts
  in KKLT},'' \href{http://arxiv.org/abs/1902.01412}{{\ttfamily
  arXiv:1902.01412 [hep-th]}}.

\bibitem{Gautason:2019jwq}
F.~F. Gautason, V.~{Van Hemelryck}, T.~{Van Riet}, and G.~Venken, ``{A 10d view
  on the KKLT AdS vacuum and uplifting},''
  \href{http://arxiv.org/abs/1902.01415}{{\ttfamily arXiv:1902.01415
  [hep-th]}}.

\bibitem{Bena:2018fqc}
I.~Bena, E.~Dudas, M.~Gra{\~n}a, and S.~L{\"u}st, ``{Uplifting Runaways},''
  {\em Fortsch. Phys.} {\bfseries 2018} (2018) 1800100,
  \href{http://arxiv.org/abs/1809.06861}{{\ttfamily arXiv:1809.06861
  [hep-th]}}.

\bibitem{Blumenhagen:2019qcg}
R.~Blumenhagen, D.~Kläwer, and L.~Schlechter, ``{Swampland Variations on a
  Theme by KKLT},'' \href{http://dx.doi.org/10.1007/JHEP05(2019)152}{{\em JHEP}
  {\bfseries 05} (2019) 152},
\href{http://arxiv.org/abs/1902.07724}{{\ttfamily arXiv:1902.07724 [hep-th]}}.
%%CITATION = ARXIV:1902.07724;%%.

\bibitem{Maldacena:2000mw}
J.~M. Maldacena and C.~Nunez, ``{Supergravity description of field theories on
  curved manifolds and a no go theorem},''
  \href{http://dx.doi.org/10.1142/S0217751X01003935;
  10.1142/S0217751X01003937}{{\em Int. J. Mod. Phys.} {\bfseries A16} (2001)
  822--855}, \href{http://arxiv.org/abs/hep-th/0007018}{{\ttfamily
  arXiv:hep-th/0007018 [hep-th]}}.

\bibitem{Hertzberg:2007wc}
M.~P. Hertzberg, S.~Kachru, W.~Taylor, and M.~Tegmark, ``{Inflationary
  Constraints on Type IIA String Theory},''
  \href{http://dx.doi.org/10.1088/1126-6708/2007/12/095}{{\em JHEP} {\bfseries
  12} (2007) 095}, \href{http://arxiv.org/abs/0711.2512}{{\ttfamily
  arXiv:0711.2512 [hep-th]}}.

\bibitem{Bergman:1999kq}
O.~Bergman and M.~R. Gaberdiel, ``{NonBPS states in heterotic type IIA
  duality},'' \href{http://dx.doi.org/10.1088/1126-6708/1999/03/013}{{\em JHEP}
  {\bfseries 03} (1999) 013},
\href{http://arxiv.org/abs/hep-th/9901014}{{\ttfamily arXiv:hep-th/9901014
  [hep-th]}}.
%%CITATION = HEP-TH/9901014;%%.

\bibitem{Witten:1998cd}
E.~Witten, ``{D-branes and K theory},''
  \href{http://dx.doi.org/10.1088/1126-6708/1998/12/019}{{\em JHEP} {\bfseries
  12} (1998) 019}, \href{http://arxiv.org/abs/hep-th/9810188}{{\ttfamily
  arXiv:hep-th/9810188 [hep-th]}}.

\bibitem{Uranga:2000xp}
A.~M. Uranga, ``{D-brane probes, RR tadpole cancellation and K theory
  charge},'' \href{http://dx.doi.org/10.1016/S0550-3213(00)00787-2}{{\em Nucl.
  Phys.} {\bfseries B598} (2001) 225--246},
  \href{http://arxiv.org/abs/hep-th/0011048}{{\ttfamily arXiv:hep-th/0011048
  [hep-th]}}.

\bibitem{Blumenhagen:2006ci}
R.~Blumenhagen, B.~K{\"o}rs, D.~L{\"u}st, and S.~Stieberger,
  ``{Four-dimensional String Compactifications with D-Branes, Orientifolds and
  Fluxes},'' \href{http://dx.doi.org/10.1016/j.physrep.2007.04.003}{{\em Phys.
  Rept.} {\bfseries 445} (2007) 1--193},
  \href{http://arxiv.org/abs/hep-th/0610327}{{\ttfamily arXiv:hep-th/0610327
  [hep-th]}}.

\bibitem{Ibanez:2012zz}
L.~E. Ib{\'a}{\~n}ez and A.~M. Uranga, {\em {String theory and particle
  physics: An introduction to string phenomenology}}.
\newblock Cambridge University Press, 2012.
\newblock
  \url{http://www.cambridge.org/de/knowledge/isbn/item6563092/?site_locale=de_DE}.

\bibitem{Sen:1999mg}
A.~Sen, ``{NonBPS states and Branes in string theory},'' in {\em {Supersymmetry
  in the theories of fields, strings and branes. Proceedings, Advanced School,
  Santiago de Compostela, Spain, July 26-31, 1999}}, pp.~187--234.
\newblock 1999.
\newblock \href{http://arxiv.org/abs/hep-th/9904207}{{\ttfamily
  arXiv:hep-th/9904207 [hep-th]}}.

\bibitem{Frau:1999qs}
M.~Frau, L.~Gallot, A.~Lerda, and P.~Strigazzi, ``{Stable nonBPS D-branes in
  type I string theory},''
  \href{http://dx.doi.org/10.1016/S0550-3213(99)00624-0}{{\em Nucl. Phys.}
  {\bfseries B564} (2000) 60--85},
\href{http://arxiv.org/abs/hep-th/9903123}{{\ttfamily arXiv:hep-th/9903123
  [hep-th]}}.
%%CITATION = HEP-TH/9903123;%%.

\bibitem{MarchesanoBuznego:2003axu}
F.~G. {Marchesano Buznego}, {\em {Intersecting D-brane models}}.
\newblock PhD thesis, Madrid, Autonoma U., 2003.
\newblock \href{http://arxiv.org/abs/hep-th/0307252}{{\ttfamily
  arXiv:hep-th/0307252 [hep-th]}}.

\bibitem{Blumenhagen:2015kja}
R.~Blumenhagen, A.~Font, M.~Fuchs, D.~Herschmann, E.~Plauschinn, Y.~Sekiguchi,
  and F.~Wolf, ``{A Flux-Scaling Scenario for High-Scale Moduli Stabilization
  in String Theory},''
  \href{http://dx.doi.org/10.1016/j.nuclphysb.2015.06.003}{{\em Nucl. Phys.}
  {\bfseries B897} (2015) 500--554},
  \href{http://arxiv.org/abs/1503.07634}{{\ttfamily arXiv:1503.07634
  [hep-th]}}.

\end{thebibliography}\endgroup
\bibliographystyle{utphys}

%%%%%%%%%%%%%%%%%%%%%%%%%%%%%%%%%%%%%%%%%%%%%%%
%%%%%%%%%%%%%%%%%%%%%%%%%%%%%%%%%%%%%%%%%%%%%%%
%%%%%%%%%%%%%%%%%%%%%%%%%%%%%%%%%%%%%%%%%%%%%%%
%%%%%%%%%%%%%%%%%%%%%%%%%%%%%%%%%%%%%%%%%%%%%%%

\end{document}